\def\year{2024}\relax
\documentclass[letterpaper]{article} % DO NOT CHANGE THIS
\usepackage[english]{babel}
\usepackage{aaai24}  % DO NOT CHANGE THIS
\usepackage{times}  % DO NOT CHANGE THIS
\usepackage{helvet}  % DO NOT CHANGE THIS
\usepackage{courier}  % DO NOT CHANGE THIS
\usepackage[hyphens]{url}  % DO NOT CHANGE THIS
\usepackage{graphicx} % DO NOT CHANGE THIS
\usepackage{natbib}  % DO NOT CHANGE THIS AND DO NOT ADD ANY OPTIONS TO IT
\usepackage{caption} % DO NOT CHANGE THIS AND DO NOT ADD ANY OPTIONS TO IT
\DeclareCaptionStyle{ruled}{labelfont=normalfont,labelsep=colon,strut=off} % DO NOT CHANGE THIS
\usepackage{algorithm}
\usepackage{algorithmic}
\usepackage{xcolor}
\usepackage{newfloat}
\usepackage{listings}
\floatstyle{ruled}
\newfloat{listing}{tb}{lst}{}
\floatname{listing}{Listing}

\usepackage{booktabs,makecell,multirow,array,colortbl,dcolumn}
\usepackage{listings}
\usepackage{multirow}
\usepackage{tikz}
\usetikzlibrary{arrows,shapes,backgrounds}
\usepackage{verbatim}
\usepackage{amsmath,amsfonts,amssymb}

\nocopyright
\title{You are a Bot! - Studying the Development of Bot Accusations on Twitter }
\author{
    Dennis Assenmacher\textsuperscript{\rm 1}\equalcontrib,
    Leon Fröhling\textsuperscript{\rm 1}\equalcontrib,
    Claudia Wagner\textsuperscript{\rm 1,}\textsuperscript{\rm 2}
}
\affiliations{
    \textsuperscript{\rm 1}GESIS - Leibniz Institute for the Social Sciences, Cologne, Germany\\
     \textsuperscript{\rm 2} RWTH Aachen, Aachen, Germany
}

\begin{document}
\tikzstyle{every picture}+=[remember picture]
\tikzstyle{na} = [baseline=-.5ex]
\maketitle

%% The abstract is a short summary of the work to be presented in the
%% article.
\begin{abstract}
 The characterization and detection of bots with their presumed ability to manipulate society on social media platforms have been subject to many research endeavors over the last decade.
In the absence of ground truth data (i.e., accounts that are labeled as bots by experts or self-declare their automated nature), researchers interested in the characterization and detection of bots may want to tap into the wisdom of the crowd. But how many people need to accuse another user as a bot before we can assume that the account is most likely automated? And more importantly, are bot accusations on social media at all a valid signal for the detection of bots? 
Our research presents the first large-scale study of bot accusations on Twitter and shows how the term bot became an instrument of dehumanization in social media conversations since it is predominantly used to deny the humanness of conversation partners. Consequently, bot accusations on social media should not be naively used as a signal to train or test bot detection models.

 %In this systematic data-driven study, we explore the users' perception of the construct bot at a large scale, focusing on the evolution of bot accusations over time. We create and analyze novel datasets consisting of bot accusations that have occurred on the social media platform Twitter since 2007, providing insights into the meanings and contexts of these particular communication situations. We find evidence that over time the term bot has moved away from its technical meaning to become an "insult" specifically used in polarizing discussions to discredit and ultimately dehumanize the opponent.
\end{abstract}

%%
%% Keywords. The author(s) should pick words that accurately describe
%% the work being presented. Separate the keywords with commas.

%%
%% This command processes the author and affiliation and title
%% information and builds the first part of the formatted document.
\maketitle

\section{Introduction}
Automated accounts on social media platforms (bots) have gained a lot of attention from a broader public in recent times, as tech entrepreneur Elon Musk, in a turbulent year for the social media platform Twitter ($\mathbb{X}$)\footnote{In this work we will refer to the platform as Twitter, which was its name at the time our data was collected.}, started bringing up his quest to combat bots among the main motivations to acquire the platform in spring 2022. He then did not agree with Twitter's estimates of the prevalence of bots on the platform and tried to use this disagreement as justification for terminating the agreed-upon deal. After ending up acquiring the platform anyway, Musk made \textit{fighting the bots} a top priority for the platform's remaining resources. While the academic discourse on bots usually receives a lot less attention than during the months of the Twitter takeover, the questions of \textit{whether}, \textit{how many}, and \textit{what types of bots} act on online discussion platforms have a tradition of being heavily discussed. The issues of who is considered a bot and what their prevalence and influence really is did not just arise in 2022, but researchers have been investigating the phenomenon of  social bots for years. Bots have been found (partly) responsible for many electoral surprises, like the outcome of the Brexit referendum \cite{howard_bots_2016} and the election of Donald Trump as U.S. president in 2016 \cite{bessi2016social}, or large-scale disinformation campaigns, for instance around many aspects of COVID-19 \cite{himelein-wachowiak_bots_2021,ferrara_what_2020}. These studies focus on identifying and measuring the influence that bots, through the propagation of certain opinions and sentiments in online networks, have on human users, allegedly allowing the bot operators to not only steer the online discourse but also impact important offline events.

While most research on social bots is focused on the development of detection methods and the characterization of suspected bot operations \cite{yan2022landscape}, in this exploratory study, we switch perspectives and approach the issue of bots on social media platforms in general and Twitter in particular through the user's perspective.
We present the first large scale study on bot accusations on Twitter and explore to what extent those accusations may be useful for training automated bot detection systems. Further, we test the assumption that "bot" is used as a pejorative term to indicate disagreement and discredit \cite{wischnewski2022agree,halperin2021bots} on a larger scale. 

%Similar to some more recent studies, we propose to explore what ordinary Twitter users think about bots on the platform, how they perceive them and how they interact with them \textcolor{blue}{through direct accusations}. However, 
Unlike previous research we do not rely on settings where participants are faced with constructed bot accounts in an experimental setup and later surveyed for their experiences \cite{yan_asymmetrical_2021, wischnewski2021disagree}, but follow an empirical approach and analyze inter-user communications, in particular situations in which one Twitter user \textit{accuses} another one of being a bot.\footnote{We refer to a \textit{user} as a Twitter account that could be potentially controlled by either a real human or an automated software.} This allows us to not only explore the characteristics of the accounts frequently accused as bots by other Twitter users, but also gives us insights into the topical contexts as well as the motivation and reasoning provided in the accusations, as they often contain justification for the verdict. Leveraging data from Twitter's inception in 2007, we explore the context and meaning of the bot accusations from different perspectives and track their evolution over the long term.
\newpage

Concretely, we follow three distinct research questions:
\begin{itemize}
\item RQ1: How did bot accusations change over time?
\item RQ2: What are the contexts in which Twitter users accuse others of being a bot? 
\item RQ3: Do the definitions of bots internalized by Twitter users align with the definitions used in popular bot detection methods?
\end{itemize}

The implications of our research are twofold. Firstly, our research has implications for scholars and practitioners interested in developing automated bot detection systems since our results show that bot accusations on social media should not be naively used as a signal to train bot detection models or as ground-truth data (which is rarely available) to test such models. Secondly, our research provides empirical support for Haslam`s theoretical argument that dehumanization is not only an important phenomenon in the intergroup context but also in the interpersonal context \cite{haslam2006}. Our results highlight that the term bot became an instrument of dehumanization since it is predominantly used to deny the humanness of conversation partners.

With this exploratory analysis, we publish the underlying "You're a bot!" datasets, containing the accusation situations that we analyze throughout the following pages.\footnote{All analysis scripts and a list of Tweet IDs can be found in the following repository: \url{https://github.com/Dennis1989/YaB}.} %For reviewers, we attach everything needed for replication of our results as supplementary material. Due to the recent API restrictions of Twitter, we will share the full data upon request.} 
These datasets allow for a number of follow-up studies, including an investigation of the campaigns and types of bots recognized by Twitter users and a more in-depth, qualitative study of the Tweet patterns and user profile features that might have triggered the bot accusation. %While this dataset contains the accounts that are recognized as and accused of being bots by other Twitter accounts, we advise against using these accounts as training or evaluation data for research developing bot detection methods. 
As we hope to motivate in the following study, we would like to see research efforts being redirected towards studying the effects of bot accusations on individual users and the impacts of the dismissal of human opinions as "bot chatter" on the public discourse.

\section{Related Work}
Only very recently, bot research has begun to move from working on detection possibilities to focusing on the user perspective, asking how platform users interact with and are affected by bots. \citeauthor{wischnewski2021disagree} study the perceptions of political social bots on Twitter in an experimental study and investigate their participants' ability to distinguish explicitly partisan bots they created from human Twitter users. Their \textit{motivated reasoning} hypothesis assumes that \textit{opinion-incongruent} accounts, i.e., accounts of different political beliefs, are more likely to be perceived as bots. While they have to reject this hypothesis for the main set of participants, they find that it holds true for the more experienced Twitter users, leading them to speculate that this effect stems from a different usage of the term (social) bot between the two groups. According to the authors, \textit{"participants with prior knowledge of social bots and participants who spend more time on social media might apply the term social bot as a pejorative term to indicate disagreement and discredit accounts"} \cite{wischnewski2021disagree}. While they speculate that the desire to "show disagreement by labeling accounts as social bots (expressive disagreement)" amplifies the effect of motivated reasoning, they do not find other evidence for this claim than a single blog post from 2019\footnote{https://saoornik.medium.com/everybody-i-dont-agree-with-is-a-russian-bot-or-how-it-is-easier-to-believe-an-evil-mastermind-ca02391055cb} and a general reference to "popular media".

In a follow-up study from the same research group, using a similar experimental design, \citeauthor{wischnewski2022agree} investigate the willingness of Twitter users to interact with \textit{social bots}. They hypothesize and subsequently show that users are more likely to interact with \textit{bots} that are more human-like, and that they are generally more likely to interact with \textit{opinion-congruent} bots, i.e., those bots perceived to be of the same partisanship as the user \cite{wischnewski2022agree}. While their hypothesis that affirmative actions (following and retweeting) are more likely to be directed towards opinion-congruent accounts holds, their assumption that \textit{motivated reasoning} does not apply to more ambiguous actions (quote-tweeting and replying) is rejected. Their hypotheses that those actions are thus equally likely to be directed towards \textit{opinion-congruent} and \textit{-incongruent} accounts do not hold. They thus find that Twitter users are more likely to interact with other accounts they perceive to be of the same partisanship, no matter whether in actions signaling agreement or being of a more ambiguous nature.

As an exception from the purely experimental studies presented before, \citeauthor{halperin2021bots} investigates the user discourse about automated manipulation on Facebook pages in the context of the lead-up to Israel's national elections in 2019 in a data-driven way. The author qualitatively assesses 525 user comments originating from Israeli politicians' posts in which bots are explicitly mentioned. He finds that users widely differ in their understanding of the construct and use it to push their political agenda. Moreover, Halperin finds that partisan commenters \textit{"strip the term of its original technical connotations and recast it as a pejorative concept used to belittle and delegitimize human right-wingers"} \cite{halperin2021bots}. Halperin concludes that this new direction in the discourse on automated accounts leads to new possibilities to attack opponents in an online conversation and may even increase the division between political camps. While the study provides interesting insights into discussions and accusations around bots, it's anecdotal nature (focus on a small sample of comments in context of a concrete political event) is mentioned as a limitation by the author, who emphasizes the need for larger studies across the social media sphere with additional textual analyses.

\citeauthor{toernberg2022} develops a model for the increased polarization of society, rejecting the echo chamber hypothesis (selective exposure and isolation from opposing views drive polarization) in favor of one based on affective polarization, where the interactions outside of the local bubble, facilitated by social media, drive polarization \cite{toernberg2022}. Affective polarization literature finds that opposing partisans have grown to "dislike, even loathe" each other \cite{iyengar2012}. The rejection of the echo chamber hypothesis by \citeauthor{toernberg2022} is supported by empirical evidence, finding that social media users of opposing political worldviews do indeed interact with each other \cite{barbera2015tweeting}. However, \citeauthor{toernberg2022} describes these interactions as "contentious and conflictual" rather than "rational arguments and deliberations" held in good faith. In conclusion, \citeauthor{toernberg2022} describes the affective polarization arising on social media as characterized by "difference, distrust, and disdain for one's political opponent".

\subsection{Bot Definitions}
Providing a clear definition of what a bot is remains one of the main challenges for any research on bots in a social media context. %The inherent problem is that there is no such thing as a comprehensive ground truth dataset of different bots from which an empirically-grounded definition could readily be derived. 
Researchers often ground their definition of the concept bot in past publications, citing a range of studies that report having found bots acting in different contexts (e.g., manipulating the discourse on a certain topic or influencing voters and impacting elections). \citeauthor{grimmesocialbots} provides an overview of the changing understandings of the concept of "bot" in academia. In the early years, bots occurred only as chat-bots, built for a specific topic and deployed in one-to-one communication settings \cite{grimmesocialbots}. This changed when spam-bots were initially developed, specialized in one-to-many-communication, and therefore key to quickly and widely amplifying content. Finally, a more recent definition that goes beyond the aspects of automation and  one-to-many-communication comes from \citeauthor{ferrara2016rise}, who define a bot as a program \textit{"that automatically produces content and interacts with humans on social media"} \cite{ferrara2016rise}. This definition is often additionally extended by the notion that these bots attempt to mimic human users \cite{abokhodair2015dissecting,stieglitz2017social}. In political contexts, a specific objective is frequently added to the definition; smearing opposing candidates \cite{metaxas2012social}, drowning political discussions \cite{thomas2012adapting}, or interfering with important political events \cite{woolley2016political,bastos2019brexit,bessi2016social}.
Once the first hurdle of providing a definition is cleared, researchers need to operationalize that definition to differentiate between bots and non-bots in a given dataset. A frequent reason for a deficit in validity in social media research is a mismatch between the definition for a specific construct (e.g., bot) and an operationalization that is measuring a different construct \cite{sen2021total}. While some researchers use account characteristics (username, profile description, profile image), activity (frequency of posting, share of retweets), or content (amount of hashtags, amount of URLs) as features to construct rules upon \cite{kollanyi2016bots}, which are then used to classify accounts as bots, many others rely on the data-driven nature of supervised Machine Learning (ML) methods. The most prominent example of this approach is \textit{Botometer} \cite{davis2016botornot}. Since its publication, \textit{Botometer} has been used to detect bots in several studies in different contexts, with its API and the corresponding ease of use contributing strongly to its popularity. Supervised ML approaches do not require hand-craft rules based on which accounts are classified but leverage the ability of ML to identify patterns in a large number of features and thereby learn classification rules directly from the provided training data. The operationalization in these approaches is then not stated explicitly but induced by the training data and represented by the trained model. This makes it even more complicated to align definition and operationalization, requiring researchers to check whether the characteristics of the instances labeled as bots in the training data are the same as those implied by definition.   

%\section{The \textit{You're a Bot} Dataset}
\section{Accusation Datasets}
\begin{figure}
\begin{center}
\includegraphics[width=0.5\textwidth]{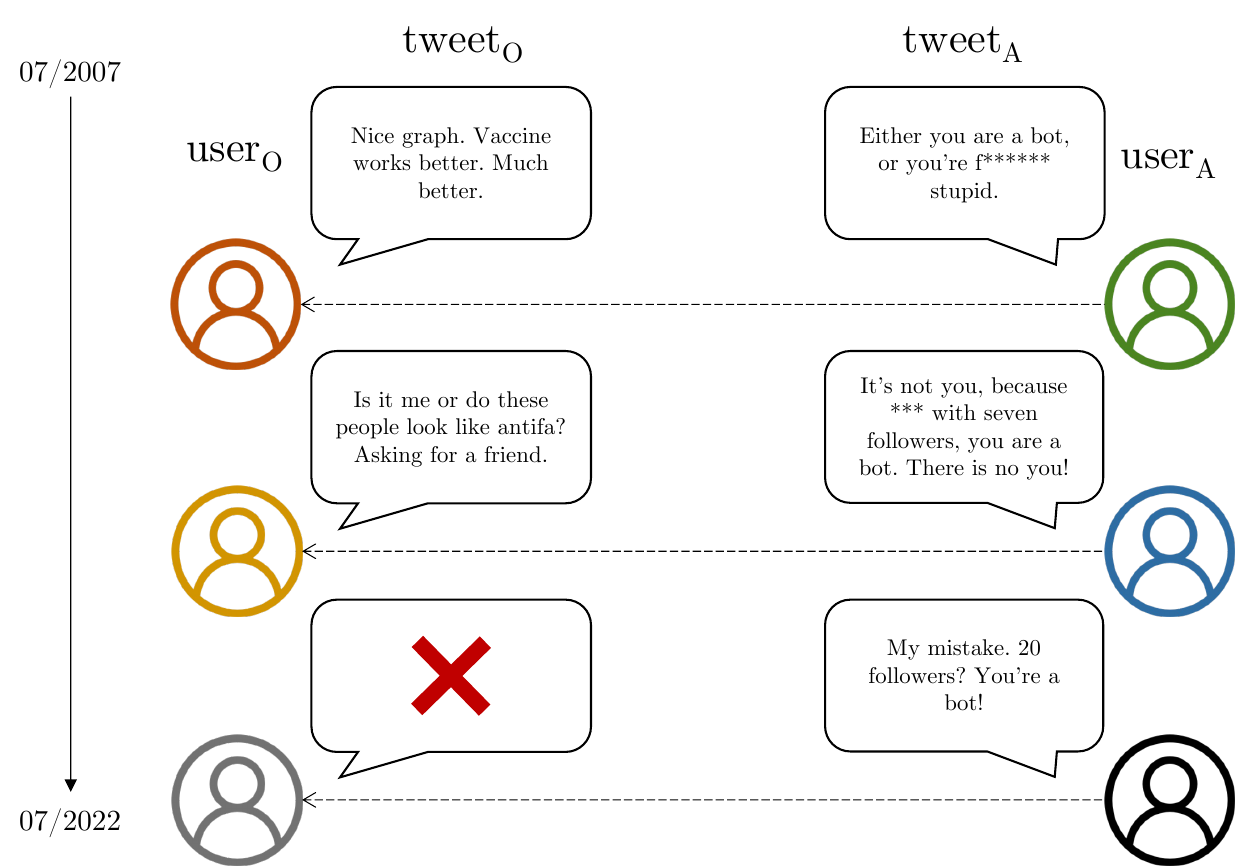}
\end{center}
\caption{Dataset structure with different accusation situations over time. Because of the data collection strategy, the objects $user_A$ and $tweet_A$ are always present in any accusation situation, whereas the objects $user_O$ and $tweet_O$ could be missing since the tweet and/or the user account have been deleted and are consequently no longer available through the Twitter API.}
\label{fig:dataset_structure}
\end{figure}

We are interested in accusation situations; communication situations in which one Twitter user accuses another user of being a bot. We collect these situations by searching for bot accusations in the set of all Reply-Tweets that contain the keyword bot. The accusation situations we collect therefore always follow the same pattern. One Twitter user ($user_O$) posts a Tweet ($tweet_O$ - we refer to this Tweet as the \textit{original} Tweet, since it is the first Tweet in our accusation situation, however, it is not necessarily the first Tweet in the full conversation) of any type or content, to which another Twitter user ($user_A$) replies with a Tweet containing a bot accusation ($tweet_A$) (see Figure \ref{fig:dataset_structure}).
To construct a dataset of situations in which one Twitter user accuses another Twitter user of \textit{being a bot}, we split the dataset creation process into two phases. During the first phase, we collect as many \textit{candidate} accusation situations as possible, accepting an increased number of false positives while aiming for a high recall. In the second phase, we filter out instances deemed irrelevant to our research design, hoping to reduce the number of false positives while ideally maintaining high precision. In the following, we describe the data collection (first phase) and data processing (second phase), the resulting datasets, and provide some descriptive statistics of them.

\subsection{Data Collection}
\label{subsection:data_collection}
For data collection, we used the academic access to the Twitter v2 API in order to obtain access to its full-archive search endpoint.\footnote{\url{https://developer.twitter.com/en/docs/twitter-api/tweets/search/api-reference/get-tweets-search-all}} This endpoint returns all Tweets matching a given query. It is important to note that the endpoint employs a token-based strategy for matching the keyword specified in the query, meaning that only Tweets containing the keyword either as a freestanding word or preceded and/or followed by a punctuation mark are matched by the API. Together with the other known but inevitable limitation of Tweets that have been removed or deleted from the platform not being available for retrospective collection through the API, this collection strategy allows us to collect all Tweets matching our query since the inception of Twitter in 2007 \footnote{Please be informed that the academic API was discontinued in April 2023.}.

We utilized the query \textit{"bot is:reply lang:en"} to match all Tweets containing the key-token "bot" that are sent as a reply to another Tweet and that are classified as written in English by the Twitter API. Our choice of query and particularly the focus on Reply-Tweets serves as the first necessary step to collect what we are calling accusation situations, i.e., situations in which one user accuses another user of \textit{being a bot}. By restricting our data collection to English Tweets, we avoid complications associated with the processing and analysis of textual data in different languages. As English is the main language used on Twitter and the default language used by international and professional audiences, we are confident that even with this restriction, we cover an interesting share of the bot accusations on Twitter. %While we might miss the properties of bot accusations in different linguistic and cultural contexts, this allows us to focus on the parts of the public discourse in which the discussion around bot accounts seems most advanced and the general public is most aware of connected issues. 
However, any findings will, by design, only apply to English-speaking user communities on Twitter.

After the data retrieval for the whole available period, beginning April 2007 and ending December 2022, %with a predefined cut-off date, July 2022, 
we were left with a dataset containing 22,275,139 Tweets\footnote{This is already filtered down from the 39,896,323 Tweets returned from the API queries. We removed Tweets that had the bot token only in the Tweet's mentions (and not in the remaining \textit{main} text), as these Tweets might be directed towards an account with \textit{bot} as part of the username without mentioning the term bot in the main message, e.g. "@bot123 how is your day?"}; replies that contained the keyword bot and that we thus consider \textit{potential} accusation situations ($bot_{all}$). As visualized in Figure \ref{fig:dataset_structure}, each accusation situation ideally consists of four different objects (and their associated metadata). However, since we are matching any potential accusation situation from the Twitter API solely based on $tweet_A$, we can only be sure that $tweet_A$ and $user_A$, i.e., the \textit{accusing} Tweet and the user who sent it, are actually available in our dataset. This is because both objects must have been available at the moment of data collection, as otherwise they would not have been matched via the API. $tweet_O$ and $user_O$, on the other hand, were only included in the dataset if they have not been deleted on Twitter and consequently were still available at the moment of data collection. Otherwise those instances have been marked as missing in our dataset.

\subsection{Data Processing}
Based on our data collection strategy, it is evident that not all of the retrieved tweets are relevant, as they may not contain bot accusations and need to be filtered accordingly. To address this issue, we decided to implement a two-phase filtering approach. In the first phase, we fine-tuned a language model for detecting bot accusations.
%on a manually annotated random subset of our data. 
Two expert annotators labelled 2,000 potential accusations that were randomly sampled from the $bot_{all}$ dataset to fine-tune the model (training details are provided in the Methods Section).
The annotation task was framed as a binary classification problem, determining if a Tweet posted by $user_A$ is \textit{"accusing another Twitter user of being a bot, irrespective of whether the accusation is implicit or explicit."}. 
By training on this randomly sampled, annotated subset, the final model learned a wide range of different accusation patterns and was able to exclude irrelevant Tweets, such as those containing self-accusations like "I am a bot." Using the fine-tuned model, we filtered the original dataset down to more than 9 million accusation situations ($bot_{general}$), which were used to investigate the evolution of accusations (RQ1) and the topical accusation contexts (RQ2). However, for the investigation of the agreement between accusations and bot scores (RQ3), we had to introduce an additional filtering step. While our model is able to identify accusations, it is often not clear if $user_A$ is truly accusing $user_O$, for example, in Tweets like \textit{"@User1 @User2 @User3 Clearly a bot! stupid moron."} Since RQ3 is based on the assumption of direct accusations, we introduced a second filtering step to exclude any ambiguous tweets.

%We were expecting the data collected in the first phase to be high in recall but low in precision (a lot of the collected reply situations would not represent the accusation situations we are interested in). 
To increase the precision of the dataset used to answer RQ3, we filtered the situations, retaining only the ones in which $tweet_A$, the \textit{accusing} Tweet, contained the regular expression \textit{"you are a [a-z]*bot$|$you're a [a-z]*bot"} ($bot_{direct}$). While this inevitably lowered the recall of our dataset by dropping situations in which the bot accusation does not explicitly follow our template (e.g., "You look like a bot to me.") or is more verbose in its use of the template (e.g., "You are a stupid bot!"), we deliberately decided to trade off some of the high recall of the $bot_{general}$ dataset for a very high precision in the $bot_{direct}$ dataset. Using this rather strict template, we are confident that the resulting dataset almost exclusively contains situations in which $user_O$ is directly accused of being a bot. %While there undeniably are some cases in which the phrase "you are a bot" is not equivalent to accusing someone of being a bot, for example if used ironically or in paraphrasing someone else, they are still part of the context and usage of the phrase and therefore worth including in our analysis.

To summarize, our prepossessing steps resulted in three datasets of different granularity that help us to investigate our research questions:
\begin{itemize}
\item $bot_{all}$ (all Reply-Tweets containing "bot")
\item $bot_{general}$ ($user_A$ accuses some other user)
\item $bot_{direct}$ ($user_A$ accuses $user_O$)
\end{itemize}

\begin{figure}
    \centering
    \includegraphics[width=0.5\textwidth]{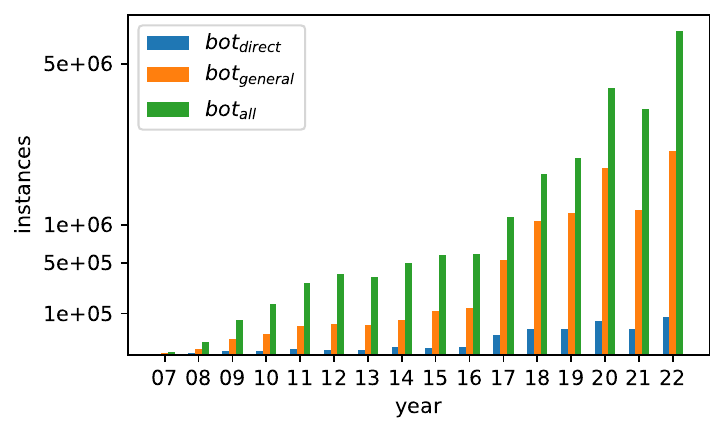}
    \caption{Number of instances in the different data subsets. While the sizes of the different datasets differ, their developments follow the same trends, like the steep increase after 2017.}
    \label{fig:accusations_over_time}
\end{figure}

The distributions of all datasets can be seen in Figure \ref{fig:accusations_over_time}.

\subsection{Dataset Characteristics}
In each of the datasets in Figure \ref{fig:accusations_over_time}, the distribution of instances over time hints at a shift that has happened in the prevalence of bot accusations around 2017. Prior to 2017, the average yearly share of $bot_{general}$ accusations in the number of $bot_{all}$ Tweets collected was at 17.4\%. In the years since 2017, the share of $bot_{general}$ accusations increased drastically, to an average yearly share of 46.3\%. 
% This is very speculative and hard to understand at that point in time.
%The fact that it is not just the absolute number of instances in the datasets that is increasing, but since 2017 also the share of accusations relative to the general chatter about bots, might already be indicative of a change in how the term "bot" as well as the bot accusation is since being used to accuse a Twitter user of \textit{being a bot}.\\

%\wac{Kann man hier eine tabelle machen? Der Teil liest sich sehr schwer wegen den ganzen Zahlen. Ich weiss auch nicht ob der Lese die infos zu den incomplete instances braucht. ich weiss nicht genau was ich mit der info tun soll und es macht ein schlechtes Gefühl weil ja soviel unvollständig zu sein scheint. Vielleicht reicht es pro dataset sowas zu haben: number of tweets, proportion of unique tweets, proportion of tweet\_a und Tweet\_o, Number of users, proportion of unique users}
As discussed above, it was not possible to collect the full accusation situation as depicted in Figure \ref{fig:dataset_structure} for all instances in the datasets due to Tweet or user account deletions. Table \ref{tab:composition} shows the composition of the two accusation datasets that we use for our analysis.
%Of the 9,069,182 accusations in the $bot_{general}$ dataset, 5,936,705 are complete, with information on $tweet_O$, $tweet_A$, $user_O$ and $user_A$ successfully collected. Of the 3,132,477 incomplete instances, 2,536,108 do not have info on $user_O$ and 3,131,278 do not have info on $tweet_O$.} Of the \textcolor{blue}{309,107} total accusation situations included in the $bot_{direct}$ dataset, \textcolor{blue}{193,623} are complete. Of the \textcolor{blue}{115,484} incomplete instances, \textcolor{blue}{96,289} lack information on $user_O$, and for \textcolor{blue}{115,464} no information on $tweet_O$ could be collected. 

\begin{table}
\caption{Composition of the different accusation datasets.}
\label{tab:composition}
\begin{center}
\begin{tabular}{ c|c|c } 
 %\hline
 &$bot_{general}$ & $bot_{direct}$ \\ \hline
  all accusation pairs & 9,069,182 & 309,107 \\ 
  complete accusation pairs & 65,5\% & 62,6\% \\ 
  unique $user_O$  & 2,683,510  & 171,215\\ 
  unique $user_A$  & 2,391,749  & 185,254 \\ 
 % $tweet_O$&cell4 & cell5 & cell6 \\ 
 %$user_O$&cell7 & 6,533,074 & cell9 \\ 
  % $tweet_a$&cell4 & cell5 & cell6 \\ 
 %$user_a$&cell7 & cell8 & cell9 \\ 
 %\hline
\end{tabular}
\end{center}
\end{table}
%While all $tweet_A$ in the datasets are unique by design, all other objects might be included multiple times in the dataset. \textcolor{blue}{In the $bot_{general}$ dataset, we collected 5,387,682 unique $tweet_O$, 2,683,510 unique $user_O$ and 2,391,749 unique $user_A$.} In the $bot_{direct}$ dataset, there are \textcolor{blue}{189,323} unique $tweet_O$, \textcolor{blue}{171,215} unique $user_O$ and \textcolor{blue}{185,254} unique $user_A$. 

In the $bot_{direct}$ dataset for which we know that it is $user_O$ who is being accused, the $user_O$ most often accused of being a bot was accused 1,321 times, by 983 different accusers. The $user_A$ responsible for the highest number of accusations has accused 747 different $user_0$, through 1,344 Tweets. 15,748 different $user_O$ have been accused by more than one $user_A$ of being a bot, being accused on average by 3.28 different users. 24,731 $user_A$ have accused more than one $user_O$ of being a bot, accusing on average 3.67 different $user_O$ of being bots.\\

\section{Methods}
In the following we present the computational mixed-methods used to filter out accusations and answer the research questions posed in the Introduction.\footnote{The evaluation scripts can be found here: \url{https://github.com/Dennis1989/YaB}.} 

\subsection{Accusation Detection with BERT}
To filter out general bot accusations from all reply Tweets, we utilized BERTweet \cite{bertweet}, a transformer-based BERT model that was pre-trained on a large corpus of English Tweets. We annotated 2,000 randomly selected instances from our $bot_{general}$ dataset using two expert annotators, achieving an inter-annotator agreement of $\kappa$ = 0.82. During the annotation process, we encountered instances that were ambiguous and challenging to classify without additional conversational context. For these instances, we mapped them to the negative class to prioritize precision over recall. We divided the annotated data into train, validation, and test sets with a split ratio of 0.8, 0.1, and 0.1, respectively. We fine-tuned multiple models on the classification task using a hyper-parameter optimization strategy that employed random search. For this purpose, we utilized the machine learning framework Ray.\footnote{\url{https://www.ray.io}} We fine-tuned 20 individual parameter constellations resulting from the optimization strategy for four epochs. Finally, we selected the best-performing model based on its ability to filter out all accusations in our validation set. The final model achieved a macro F1 score of 0.93, recall of 0.92, and precision of 0.94 on our hold out test data.

\subsection{Word Embeddings over Time}
\label{subsection:embeddings}
Word-embeddings capture semantic similarity via word co-occurrences in vector space. Similar to previous work on shifts in the meaning of individual words over time \cite{word_embedding_shift_100y}, we track the evolution of bot accusations by training Word2Vec-embeddings on the $tweet_A$ in the $bot_{general}$ dataset collected for each year between 2007 and 2022 and inspecting the words that are most closely associated with the term \textit{bot}. We approximate the distance between words through their cosine-similarity. We use Gensim's implementation of Word2Vec with negative sampling and CBOW (as we are interested in the evolution of a high-frequency word).\footnote{\url{https://radimrehurek.com/gensim/models/word2vec.html}} To account for the well-known nondeterministic behavior of embedding models \cite{hellrich-hahn-2016-bad} and increase the robustness of our findings, we train five different embeddings for each year and consider only \textit{stable} neighbors, i.e. those words that are closely associated with the term \textit{bot} in all five embedding spaces. As the first years of the data collection consist of considerably lower numbers of Tweets (see Figure \ref{fig:accusations_over_time}), we aggregate the years from 2007 to 2016 into one collection and treat each following year individually.

\subsection{Clustering}
\label{subsection:clustering}
To understand in which topical contexts users on Twitter are accused of being bots, we identify prominent clusters in the accused users' Tweets in the $bot_{general}$ dataset through unsupervised learning. We transform their Tweets ($tweet_O$) into document embeddings using sentence transformers \cite{reimers-gurevych-2019-sentence}. Similar to the embedding approach, we use cosine-similarity to approximate distance in the resulting vector space and to group together documents in close proximity. We tested several thresholds and found that a threshold value of 0.7 resulted in the most coherent and pure clusters. The resulting clusters are projected into two-dimensional space using Uniform Manifold Approximation and Projection (UMAP) \cite{mcinnes2018umap}. For each identified cluster, we calculate a class-based variant of TFIDF (cTFIDF), which aggregates all cluster documents into one artificial document. This ensures that only the important tokens characterizing each cluster are highlighted.

\subsection{Toxicity}
To measure the toxicity of Tweets over time, we utilize the pretrained Detoxify classifier \cite{Detoxify}. The model offers a fine-grained labeling schema and is able to classify Tweets at scale. We use the pretrained base models without fine-tuning. To test the general performance of the classifier in detecting toxicity in Tweets in our specific scenario, we annotated 100 random Tweets from our $bot_{general}$ dataset manually. On this test set, Detoxify reached a balanced accuracy of 0.90. To offer an approximation of the overall toxicity on Twitter as a baseline, we collect a dataset of generic English-language replies with the same settings described in our data collection, using the query \textit{"a is:reply lang:en"}. This baseline, however, is limited to showing overall trends and patterns in the toxicity on Twitter, and we do not use it to compare absolute levels of toxicity. 

\subsection{Bot Scores}
\label{subsection:botscores}
We utilize the latest \textit{Botometer} version \cite{sayyadiharikandeh_detection_2020} to see how far the definitions of bots internalized by Twitter users, as operationalized through their accusations, align with the definition operationalized in this popular tool for bot detection on Twitter. Botometer is a supervised ML model, with the latest version using an ensemble of specialized Random Forest classifiers that were trained on a plethora of existing \textit{ground truth} bot datasets (different annotation strategies were used to identify bots in these datasets, such as honeypot traps, manual annotation, or self-report). Apart from calculating an overall score indicating how "bot-like" an account is, the classifier also returns multiple subscores corresponding to more specific bot definitions, such as "spammer" or "fake follower". Although the Botometer classifier is a helpful tool for understanding the degree of automation of a Twitter account, it comes with significant limitations. Several studies have previously demonstrated the limits of such classifiers due to concept drift in data and the uncertainty of the underlying ground truth \cite{martini2021bot, rauchfleisch2020false}. For our experiments, we utilize the method as a means to measure how the understanding of the concept bot held by Twitter users aligns with the academic construct definition operationalized by the classifier. We do not use the tool to determine how many and which accounts are bots and thus do not intend to construct a labeled ground truth dataset from our data. However, we extract a subset of accounts that are repeatedly accused of being bots by different Twitter users, for which we then collect their Botometer scores. To increase validity and tackle noise, we collect scores only for the 13,638 $user_O$ in the $bot_{direct}$ dataset that have been accused of being bots by at least two \textit{different} $user_A$. We collect their \textit{bot}-scores from the Botometer API\footnote{\url{https://botometer.osome.iu.edu/api}}.

\subsection{Ideology} % check dataset creation process + size etc.
We measure the ideology of accounts by utilizing \citeauthor{barbera2015birds}'s method for ideal point estimation \cite{barbera2015birds}. The method infers ideology by examining the political following network of Twitter accounts, assuming that social networks are homophilic. Since the method was developed using a recent set of political accounts in the United States, we restrict our analysis to users from there and to accusations made after 2016. We thus only consider accusation instances in which both $user_O$ and $user_A$ self-report as being from the United States. We use the meta-information \textit{location}, a free-text field used by some Twitter users to indicate their location, in combination with Google's Geolocation API\footnote{\url{https://developers.google.com/maps/documentation/geocoding/overview}} to automatically parse the location information and retrieve the corresponding country. 76,855 of the accusation situations in our $bot_{direct}$ dataset posted after 2016 had \textit{location}-information for both users. To handle API restrictions, we randomly extract an approximate 10\% sample, leaving us with 7,131 pairs of $user_O$ and $user_A$ for which we parse the location information with the Geolocation API. In this sample, we found 2,670 conversation pairs where both accounts are from the US, which we use as input for our ideology estimation method.

\section{Results}

In this section we answer the following three research questions:
\begin{itemize}
    \item RQ1: How did bot accusations change over time?
    \item What are the contexts in which Twitter users accuse
others of being a bot?
\item RQ3: Do the definitions of bots internalized by Twitter
users align with the definitions used in popular bot detection methods?
\end{itemize}

\subsection{RQ1: Evolution of accusations}
\label{subsection:rq1}

Table \ref{fig:embeddings} shows the nearest neighbor embeddings of the term bot over the years covered in our dataset. As the number of accusations for the early years up to 2016 is significantly lower (see Figure \ref{fig:accusations_over_time}), we aggregate the accusations made in those years to get a sufficient number of observations for training the embeddings. The color coding we applied to the nearest embeddings vectors displayed in Table \ref{fig:embeddings} shows how the meaning of the bot accusations shifted over the years. In the years leading up to 2016, the term bot is closely and firmly associated with terms representing the automated behavior often used in academia to conceptualize bots. These include \textit{software}, \textit{script}, or \textit{comments}. In the following years, the wording of the bot accusations shifted away from such signs of automation in favor of terms used to insult the accused user. 
The use of derogatory terms such as \textit{moron}, \textit{stupid}, \textit{idiot}, and \textit{shill} implies that the accuser views the accused as less than human, often by questioning their mental capacity. 
This indicates that over time bot accusations became predominantly instances of ``dehumanization'' \cite{haslam}. By studying inherently interpersonal accusation situations, we find empirical support for Haslam`s theoretical argument that ``dehumanization is an important phenomenon in interpersonal as well as intergroup contexts'' and also ``occurs outside the domains of violence and conflict'' \cite{haslam2006}.
%According to \cite{fiske1991,haslam2006,} dehumanizing conversation partners leads to ``asocial'' and ``null'' interactions, in which people ``disregard the existence of other people as social partners'' and deny the autonomous agency of other people. 

%"Dehumanizing conversation partners involves emotional distancing and represents the other as cold, robotic, passive, and lacking in depth, it implies indifference rather than disgust. Typically, mechanistically dehumanized others are seen as lacking the sort of autonomous agency that provokes strong emotion and are more likely to be seen as emotionally inert.

%Dehumanizing conversation partners creates situation where the accused is not allowed to express their opinions freely
%This explicit form of dehumanization is recognized by \citeauthor{haslam}. As a result, the accuser insults the conversation partner and denies their intelligence, creating a situation where the accused is not allowed to express their opinions freely. All of these factors contribute to a breakdown in communication and an unhealthy discourse.}
\begin{table}[]
\footnotesize
\caption{Nearest embedding vectors to the term \textit{bot} over the years. We highlight terms associated with mechanics for automation in blue and dehumanizing/insulting/political terms in red. }\begin{tabular}{ll}

2007 -&{\color{blue}\colorbox{blue!20}{\vphantom{pd}software}}, {\color{blue}\colorbox{blue!20}{\vphantom{pd}comments}}, person, user, {\color{blue}\colorbox{blue!20}{\vphantom{pd}hashtags}}, \\ 
2016&{\color{blue}\colorbox{blue!20}{\vphantom{pd}script}}, {\color{blue}\colorbox{blue!20}{\vphantom{pd}machine}}, database, guy, {\color{blue}\colorbox{blue!20}{\vphantom{pd}acct}}, {\color{blue}\colorbox{blue!20}{\vphantom{pd}program}},  \\ 
 \hline2017&{\color{red}\colorbox{pink}{\vphantom{pd}troll}}, {\color{red}\colorbox{pink}{\vphantom{pd}idiot}}, person, probably, {\color{red}\colorbox{pink}{\vphantom{pd}moron}}, \\ 
 &entity, tool, real, {\color{blue}\colorbox{blue!20}{\vphantom{pd}account}}, human, paid,  \\ 
 \hline2018&{\color{red}\colorbox{pink}{\vphantom{pd}troll}}, {\color{red}\colorbox{pink}{\vphantom{pd}idiot}}, person, probably, {\color{red}\colorbox{pink}{\vphantom{pd}moron}}, \\ 
 &{\color{red}\colorbox{pink}{\vphantom{pd}supporter}}, entity, {\color{red}\colorbox{pink}{\vphantom{pd}dolt}}, joke, {\color{red}\colorbox{pink}{\vphantom{pd}shill}}, tool, \\ 
 &writer, {\color{blue}\colorbox{blue!20}{\vphantom{pd}account}}, {\color{red}\colorbox{pink}{\vphantom{pd}fool}}, human, {\color{red}\colorbox{pink}{\vphantom{pd}russian}}, paid,  \\ 
 \hline2019&definitely, {\color{red}\colorbox{pink}{\vphantom{pd}troll}}, {\color{red}\colorbox{pink}{\vphantom{pd}idiot}}, product, {\color{red}\colorbox{pink}{\vphantom{pd}supporter}}, \\ 
 &{\color{red}\colorbox{pink}{\vphantom{pd}fool}}, human, {\color{blue}\colorbox{blue!20}{\vphantom{pd}account}}, probably, person, {\color{red}\colorbox{pink}{\vphantom{pd}moron}}, \\ 
 &parody, caricature, asset, {\color{red}\colorbox{pink}{\vphantom{pd}stooge}}, robot, joke, \\ 
 &tool, {\color{red}\colorbox{pink}{\vphantom{pd}russian}},  \\ 
 \hline2020&{\color{red}\colorbox{pink}{\vphantom{pd}troll}}, {\color{red}\colorbox{pink}{\vphantom{pd}idiot}}, {\color{red}\colorbox{pink}{\vphantom{pd}trumper}}, person, {\color{red}\colorbox{pink}{\vphantom{pd}foreigner}}, \\ 
 &teenager, probably, {\color{red}\colorbox{pink}{\vphantom{pd}supporter}}, parody, {\color{red}\colorbox{pink}{\vphantom{pd}moron}}, \\ 
 &robot, joke, tool, {\color{red}\colorbox{pink}{\vphantom{pd}fool}}, human, {\color{red}\colorbox{pink}{\vphantom{pd}russian}},  \\ 
 \hline2021&definitely, {\color{red}\colorbox{pink}{\vphantom{pd}troll}}, {\color{red}\colorbox{pink}{\vphantom{pd}idiot}}, {\color{red}\colorbox{pink}{\vphantom{pd}foreigner}}, {\color{red}\colorbox{pink}{\vphantom{pd}fool}}, \\ 
 &human, {\color{blue}\colorbox{blue!20}{\vphantom{pd}account}}, probably, {\color{red}\colorbox{pink}{\vphantom{pd}shill}}, person, parody, \\ 
 &{\color{red}\colorbox{pink}{\vphantom{pd}moron}}, paid, robot, joke, tool, real, \\ 
 &{\color{red}\colorbox{pink}{\vphantom{pd}chinese}}, {\color{red}\colorbox{pink}{\vphantom{pd}russian}},  \\ 
 \hline2022&{\color{red}\colorbox{pink}{\vphantom{pd}troll}}, {\color{red}\colorbox{pink}{\vphantom{pd}propagandist}}, {\color{red}\colorbox{pink}{\vphantom{pd}idiot}}, operative, {\color{red}\colorbox{pink}{\vphantom{pd}foreigner}}, \\ 
 &{\color{red}\colorbox{pink}{\vphantom{pd}fool}}, {\color{blue}\colorbox{blue!20}{\vphantom{pd}account}}, {\color{red}\colorbox{pink}{\vphantom{pd}simpleton}}, probably, {\color{red}\colorbox{pink}{\vphantom{pd}shill}}, \\ 
 &{\color{red}\colorbox{pink}{\vphantom{pd}liar}}, person, parody, {\color{red}\colorbox{pink}{\vphantom{pd}moron}}, satire,  {\color{red}\colorbox{pink}{\vphantom{pd}chinese}},  \\ 
 \hline\end{tabular}
\label{fig:embeddings}
\vspace{-1em}
\end{table}

\begin{figure}[t]
    \centering
    \includegraphics[width = 0.47\textwidth]{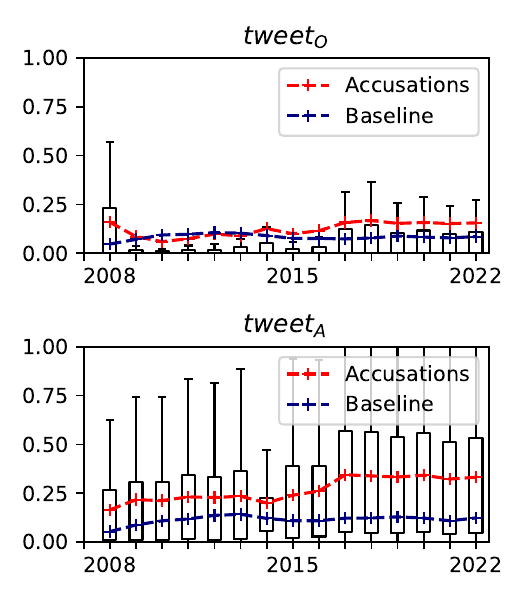}
    \caption{Development of \textit{Detoxify} toxicity scores for $tweet_O$ and $tweet_A$ objects. Comparison with the baseline of generic Reply-Tweets indicates that the observed increase in toxicity for the accusations ($tweet_A$) after 2016 cannot be explained through an overall increase of toxicity on Twitter over time.}
    \label{fig:toxicity}
\end{figure}

%\textcolor{blue}{In addition to the points made previously, it is worth noting that the use of derogatory terms to suggest low mental capabilities is particularly striking given the current state of research on social bots. Despite their assumed sophisticated ability to imitate human behavior, social media users often perceive bots as unintelligent and lacking in mental capabilities. This contradiction between the actual sophistication of bots and the public perception of them may contribute to further breakdowns in communication and misunderstandings online.}
%Terms like \textit{moron}, \textit{stupid}, \textit{idiot}, and \textit{shill} dehumanize and insult the conversation partner, question and deny their intelligence, and in last consequence, deny them their right to hold and communicate their own opinions. %Whereas the accusations in the first years of the data discuss Bots from a technical point of view, in line with the \textit{automated behavior} definition of bots often used in academia, the accusations of the later years seem to lose their grounding in a technical definition and are with terms dehumanizing and delegitimizing the conversation partner.

For direct accusations to $user_O$ ($bot_{direct}$) we observe an increase in the degree of toxicity over time. Figure \ref{fig:toxicity} shows the toxicity for $tweet_O$ and $tweet_A$ over time, as well as a baseline that serves as a proxy for how the overall toxicity in Tweet-Reply situations developed in parallel. It is evident that beginning with the year 2015, the toxicity of the $tweet_A$ increased strongly, settling at a new plateau around the year 2017. The baseline shows that this increase can not just be explained by the rise of the overall toxicity on Twitter, as the toxicity for replies outside our bot accusation context remains stable and relatively low. Similarly, the toxicity of the original Tweets leading up to the accusations is comparatively low and has a much less pronounced increase around 2016.

\subsection{RQ2: Accusation context}

The shift towards dehumanization that we observed when analyzing the bot accusations over time also becomes visible in the results of our  topical analyses. A manual investigation of the cluster solutions produced by the cluster algorithm discussed in our method section for the early years (2007-2016) revealed that users were frequently accused because they showed signs of automation, such as spamming repetitive content, for instance: 
\begin{quote} 
\textit{"follow me i follow you"}\\ 
\textit{"good morning! you deserve a fantastic day today!"}\\ 
\textit{"i'm a human i'm a human i'm a human"}.
\end{quote} 
Additionally, corporate accounts that automatically respond to public customer complaints with standardized replies like
\begin{quote}
\textit{"please dm your concern to help us assist you."}
\end{quote}
were frequently accused of being bots. We also found a considerable number of accounts that were accused because they openly discussed that they reached the limit of accounts they could follow or ran into rate limits for sending out new Tweets, such as
\begin{quote}
\textit{ oh dear, twitter says i'm not allowed to follow anymore people. what's that all about then?}.
\end{quote}

\begin{figure*}
    \centering
      \includegraphics[width=0.9\textwidth]{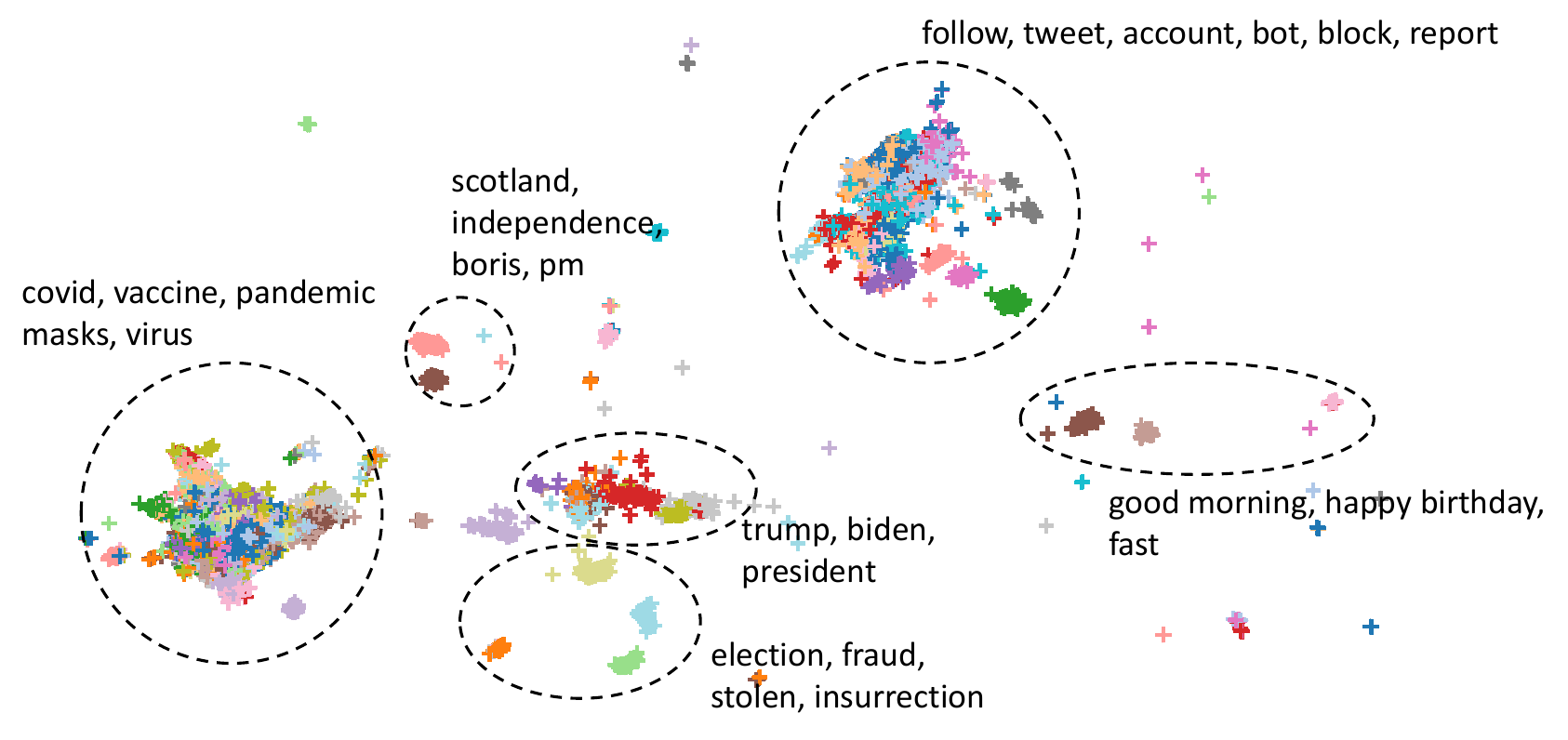}
    \caption{Top 100 clusters for the Tweets sent by $user_O$ (the user accused of being a bot) for 2021 projected onto a 2-dimensional space with UMAP. Clusters are annotated with their highest cTFIDF terms. Users are not anymore accused only in the context of automated behavior ("good morning spam") but specifically in the context of polarizing debates around topics like covid/mask/vaccine, election/Biden/Trump or Scottish independence. Additionally, we observe a large cluster indicating bot accusation loops (e.g. \textit{$user_O$: you are a bot - blocked! $user_A$: i'm not a bot, but you are!} ).} 
    \label{fig:clustering}
\end{figure*}

\begin{figure}
    \centering
    \includegraphics[width = 0.4\textwidth]{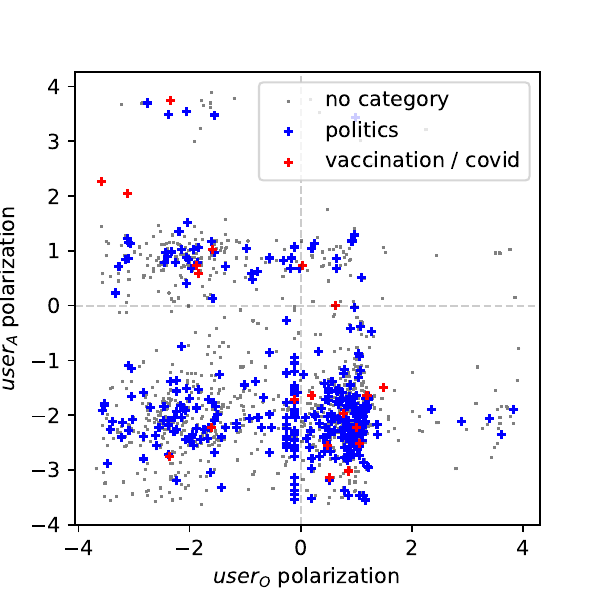}
    \caption{Ideology scores for $user_O$ and $user_A$ according to \citeauthor{barbera2015birds}\cite{barbera2015birds}. Negative scores are associated with Left-leaning political positions, while positive scores are associated with Right-leaning political positions. %The vertical axis indicates the ideology of the accusing user, the horizontal axis the ideology of the user that is being accused.
    One can see that most accusations occur in polarizing topics and are directed from left-leaning towards right-leaning users (lower right quadrant). Users on the political Right tend to not accuse each other, while users on the political Left accuse each other more frequently.}
    \label{fig:ideology}
\end{figure}

The tweet content posted by different $user_O$ that had led to an accusation situation has significantly changed over the years. Recently (especially since 2020), most $user_O$ are accused of being bots in the context of controversial, polarizing political or politicized topics such as elections (Biden vs. Trump), the pandemic (vaccination, mask mandates), or the Brexit vote (Boris Johnson). Figure \ref{fig:clustering} displays the top cluster distribution for the year 2021. Of the twenty biggest clusters (in terms of cluster size) in 2021, thirteen were concerned with political debates, and only the remaining seven focused on repetitive automation behavior (most of them indicating accusation loops). We found that specifically $user_O$ and $user_A$ communication pairs that directly accuse each other in the context of these political topics are found on opposing sides of the ideological spectrum, e.g., users from the left of the ideological spectrum accusing users from the right of the ideological spectrum in the context of the topic of Trump vs. Biden (see colored observations in Figure \ref{fig:ideology}). Interestingly, most accusers are more left-leaning, while accounts that are accused are more right-leaning (most observations are in the bottom-right quadrant in Figure \ref{fig:ideology}). When it comes to intra-ideology conversation pairs, it is evident that accounts on the political Right tend to not accuse each other (upper right quadrant), while accounts that are associated with the political Left accuse each other more frequently (lower left quadrant), especially in the context of political debates.

\subsection{RQ3: Accusations vs. Bot Scores}
\begin{figure}
    \centering
    \includegraphics[width = 0.44\textwidth]{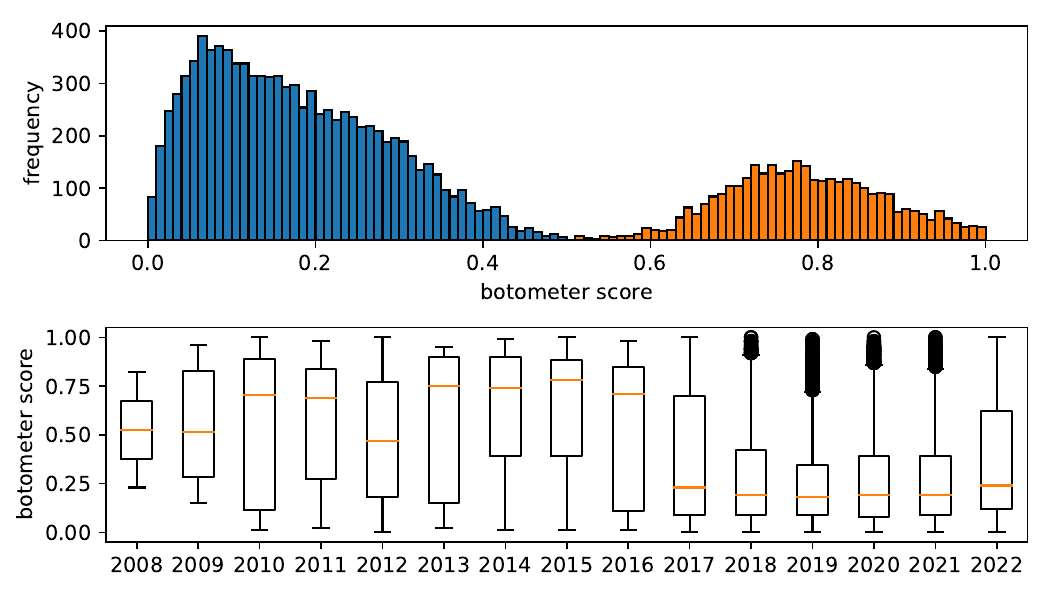}
    \caption{Distribution and development over time of Botometer scores for $user_O$ (users \textit{accused} of being bots). Between 2008 and 2016 accusations exhibited high Bot scores, while this abruptly changed beginning of 2017.}
    \label{fig:bot-scores}
\end{figure}

We investigate the alignment between bot accusations and the bot probabilities assigned by popular detection mechanisms by inspecting the Botometer scores for the $user_O$ in our $bot_{direct}$ dataset.
Figure \ref{fig:bot-scores} displays the distribution of the bot probabilities for the accused accounts as calculated by the Botometer model. The scores follow a bimodal distribution, indicating that accounts are quite confidently assigned into one of the two classes \textit{human} or \textit{bot}, with a larger fraction of accounts having lower scores and thus tending towards the \textit{human} category. Consequently, there seems to be a discrepancy between the operationalization of the bot construct employed by Botometer and the definition internalized by the Twitter users. In a subsequent analysis, we explored how bot scores developed over time. The results are depicted in the bottom part of Figure \ref{fig:bot-scores}. Interestingly, the accounts accused between 2008 and 2016 exhibit significantly higher bot scores than those between 2017 and 2022. This decrease in bot scores is well in line with the shift in the meaning of the accusations that we found before, with accusations not being exclusively used for actual automation behavior anymore. Apparently, in the early years, the Botometer scores were aligned with the accusations made by the Twitter users, as the accused users, $user_O$, also have high Botometer scores. These were the years in which the majority of $user_O$ were accused because of their automation behavior (repetitive tasks such as (re-)tweeting or large-scale following). With the observed transformation of bot accusations into a dehumanizing insult (starting in 2017), the average bot scores dropped significantly. While this is not necessarily an indicator of fewer bot occurrences (which might also be due to potential model errors), it is clearly showing that bot accusations experienced a concept drift. 

\begin{figure}
    \centering
    \includegraphics[width = 0.5\textwidth]{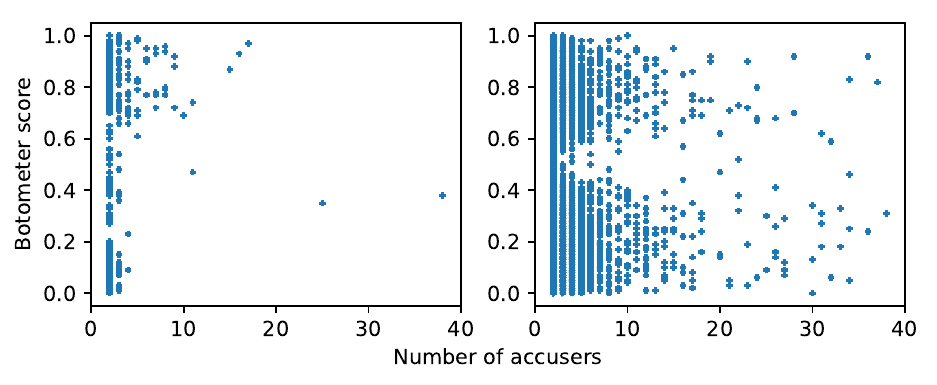}
    \caption{Scatter plot showing the relationship between the number of accusers and the Botometer score for the accused users. While high Botometer scores were associated with higher numbers of accusations that an user received before 2017 (left subfigure), this association does not hold anymore after 2017 (right subfigure).}
    \label{fig:accusations-correlation}
\end{figure}

Additionally, we investigated how Botometer's bot scores are correlated to the number of unique accusers for each $user_O$, expecting to find that the number of accusers increases with the bot score if the definitions internalized by the accusing users are well aligned with the definition operationalized in Botometer. Similar to our previous analyses, we again differentiate between the years before 2017 and from 2017 onwards. Calculating Spearman's rank correlation coefficient, we find a negligible positive correlation for the years after 2017 ($\rho=0.08,\text{p-value}<0.0001)$ and only a weak positive correlation for the early years before 2017 ($\rho=0.23,\text{p-value}=0.0001)$. However, a visual inspection of the scatter plots in Figure \ref{fig:accusations-correlation} clearly reveals that Botometer assigns only small bot scores for accounts with few accusations during the early years, whereas for all $user_O$ with more than three accusations, the scores tend to be in the higher regions (between 0.6 and 1). On the contrary, there is no clear pattern for the years from 2017 onwards (right sub-plot). Even with an increasing number of accusations, both low and high Botometer scores are present, indicating that the hypothesized and limitedly shown correlation between both variables does not hold anymore during the later years. 

\section{Discussion \& Limitations}

Our results indicate that bot accusations on the Twitter platform significantly changed over the last decade (RQ 1). Before 2017, platform accounts were mainly accused when they exhibited some explicit automation behavior. Over time and since the aftermath of the US elections in 2016, resulting in an increased interest for bots in media and academia, users increasingly accused other accounts, intending to dehumanize and insult them, questioning their intelligence and denying their right to express their opinion. In accordance with this observation, we find that bot accusations from 2017 onwards primarily occur in the context of controversial, polarizing debates like elections or covid/vaccination (RQ 2). In other words, the term "bot" has transformed into a political term that is used as an insult (or more precisely an instance of dehumanization). This starkly contrasts with the ways in which bot accounts are currently discussed in academia, focussing more on the means of automation to distribute (often) malicious content and their supposed impact on the public discourse rather than the actual usage and meaning of the term bot, as exhibited in our bot accusations. Indeed, we found that in the early years, the bot scores (calculated by Botometer) of accused accounts were significantly higher than in later years, when accusations mainly occurred as an insult to dehumanize other "people" in the network. This finding supports the theoretical argument that dehumanization is an important phenomenon
in interpersonal context \cite{haslam2006} and provides empirical evidence for the initial assumptions that "bot" is used as a pejorative term to indicate disagreement and discredit \cite{wischnewski2022agree,halperin2021bots} on a larger scale. 
Our findings also have practical implications for researchers interested in bot detection since we show that bot accusations should not be naively used as a signal for automated bot detection methods or as ground-truth data.

In addition to the investigation of how automated accounts influence users on social media platforms, future research should be concerned with the impact these accusations have on individuals. Additionally, it should be investigated how bot accusations could be countered efficiently. The structure of our dataset paves the way for a plethora of follow-up research in these directions. With additional data collection efforts, the accusation situations examined in this work could be augmented to cover the whole conversation around the bot accusations, including the exchange leading up to it, as well as the reaction afterward. 

Our study is not without limitations. Even though we tried to balance the issues associated with precision and recall by using different data collection strategies, 
%datasets, allowing us to control whether we need a very precisely directed operationalization of accusation ($bot_{direct}$) or whether the understanding of accusation could be taken much more broadly ($bot_{general}$), 
the inherent problem of not being able to achieve perfect performance in selecting accusations remains. With our data collection strategies we might, for example, have missed bot accusations that refer to bots using a synonym or the same word in a different language. %As we publish the IDs of all Reply-Tweets mentioning the term "bot", we encourage researchers to replicate our analyses using different strategies for selecting accusations. 
Also, we highlight that our ideology analysis only considered accusation situations in which both users were located in the United States. In order to use the method on accounts outside the US, the model must be trained again on a new initial set of accounts. While our methodology can be adapted to other platforms and languages, our empirical results are limited to one platform (Twitter) and one language (English). Future endeavors will show whether our findings also apply to bot accusations made on different platforms and in other linguistic and cultural settings.

\section{Ethics Statement} %and Broader Impact
This study uses publicly-accessible user-generated content online as its data source. Using this type of data carries privacy risks. The data potentially contains \textit{false accusations}, situations in which actual human users are accused of being a bot. We stress that these accusations represent subjective opinions and should under no circumstances be used to infer the degree of automation of any account. This is especially true in light of our findings that there currently is a discrepancy between the construct definition and the Twitter users' understanding of it, and that these accusations tend to happen in an increasingly toxic and polarized environment. To address these concerns, only the IDs of the Tweet- and User-objects used in this study will be publicly released. To still allow for complete reproducibility, we invite other researchers to contact us for collaboration.%share the full dataset with researchers that sign a non-disclosure agreement.

%With our new evidence that bot accusations have profoundly changed and are nowadays used in a pejorative way to insult and discredit other social media accounts, we pave the way to new social science studies that investigate the impact of these offenses on humans. Our study shows that there is a need to explore this research gap further, investigating the impact that bots and conversations around bots have on humans, moving on from focusing only on the characterization and detection of automated accounts. How do bot accusations influence the public discourse, and how do they impact concepts like trust and critical engagement that are fundamental to a functioning discourse online? Are human users offended, and how do they react to accusations? These are just a few directions that should be investigated in the future.
\section{Acknowledgments}
This work was created in context of the project: Dehumanization Online: Measurement and Consequences (DeHum), funded by the Leibniz-Gemeinschaft.

\bibliography{references} 
%instead or the References section will not appear in your paper
%\nobibliography{aaai22}

\bigskip

\end{document}